\documentclass[aps,floatfix,prb,twoside,twocolumn,10pt,showpacs,citeautoscript,superscriptaddress]{revtex4-2}

\usepackage{amsmath}
\usepackage{amssymb}
\usepackage{glossaries}
\usepackage{graphicx}
\usepackage{units}
\usepackage[version=3]{mhchem}

\usepackage[    unicode,    colorlinks=true,    urlcolor=black,    linkcolor=black,    citecolor=black]{hyperref}

\newacronym{alda}{ALDA}{adiabatic local-density approximation}
\newacronym{ar}{AR}{aspect ratio}
\newacronym{bic}{BIC}{Bayesian information criterion}
\newacronym{bz}{BZ}{Brillouin zone}
\newacronym{df}{DF}{dielectric function}
\newacronym{dft}{DFT}{density-functional theory}
\newacronym{doe}{DOE}{Department of Energy}
\newacronym{dos}{DOS}{density of states}
\newacronym{fcc}{FCC}{face-centered cubic}
\newacronym{fdtd}{FDTD}{finite-difference time-domain}
\newacronym{fwhm}{FWHM}{full width at half maximum}
\newacronym{ibt}{IBT}{interband transition}
\newacronym{lrtddft}{LR-TDDFT}{linear-response time-dependent density-functional theory}
\newacronym{lspr}{LSPR}{localized surface plasmon resonance}
\newacronym{np}{NP}{nanoparticle}
\newacronym{omp}{OMP}{orthogonal matching pursuit}
\newacronym{pml}{PML}{perfectly matched layer}
\newacronym{rmse}{RMSE}{root mean squared error}
\newacronym{rpa}{RPA}{random phase approximation}
\newacronym{sos}{SOS}{special ordered structure}
\newacronym{sqs}{SQS}{special quasi-random structure}
\newacronym{si}{SI}{Supporting Information}
\newacronym{tddft}{TDDFT}{time-dependent density-functional theory}

\begin{document}

\title{
    Computational Design of Alloy Nanostructures for Optical Sensing of Hydrogen
}

\newcommand{\phys}{
    Department of Physics,
    Chalmers University of Technology,
    SE-412~96 Gothenburg, Sweden
}
\newcommand{\aalto}{
    Department of Applied Physics,
    Aalto University,
    FI-00076 Aalto, Finland
}
\newcommand{\warsaw}{
    Faculty of Physics,
    University of Warsaw,
    Pasteura 5,
    PL-02-093 Warsaw, Poland
}

\author{Pernilla\ Ekborg-Tanner}
\author{J.\ Magnus\ Rahm}
\author{Victor\ Rosendal}
\affiliation{\phys}
\author{Maria Bancerek}
\affiliation{\warsaw}
\affiliation{\phys}
\author{Tuomas\ P.\ Rossi}
\affiliation{\aalto}
\author{Tomasz\ J.\ Antosiewicz}
\affiliation{\warsaw}
\affiliation{\phys}
\author{Paul\ Erhart}
\email{erhart@chalmers.se}
\affiliation{\phys}

\begin{abstract}

Pd nanoalloys show great potential as hysteresis-free, reliable hydrogen sensors.
Here, a multi-scale modeling approach is employed to determine optimal conditions for optical hydrogen sensing using the Pd--Au--H system.
Changes in hydrogen pressure translate to changes in hydrogen content and eventually the optical spectrum.
At the single particle level, the shift of the plasmon peak position with hydrogen concentration (\textit{i.e.}, the ``optical'' sensitivity) is approximately constant at \unit[180]{nm/$c_\text{H}$} for nanodisk diameters $\gtrsim\!\unit[100]{nm}$.
For smaller particles, the optical sensitivity is negative and increases with decreasing diameter, due to the emergence of a second peak originating from coupling between a localized surface plasmon and interband transitions.
In addition to tracking peak position, the onset of extinction as well as extinction at fixed wavelengths is considered.
We carefully compare the simulation results with experimental data and assess the potential sources for discrepancies.
Invariably, the results suggest that there is an upper bound for the optical sensitivity that cannot be overcome by engineering composition and/or geometry.
While the alloy composition has a limited impact on optical sensitivity, it can strongly affect H uptake and consequently the ``thermodynamic'' sensitivity and the detection limit.
Here, it is shown how the latter can be improved by compositional engineering and even substantially enhanced via the formation of an ordered phase that can be synthesized at higher hydrogen partial pressures.

\begin{description}
\item[Keywords] 
hydrogen sensing, nanoplasmonics, localized surface plasmon resonance, nanoparticles, palladium alloys, dielectric function
\end{description}
\end{abstract}

\maketitle

\section*{Introduction}\label{sect:introduction}

One of the main challenges of the hydrogen economy, a future energy system where fossil fuels have been replaced with hydrogen-based fuels, is the flammability of hydrogen gas under ambient conditions \cite{Dut14}.
As a result, hydrogen sensing continues to be a very active research field with the goal of providing fast, reliable, and long-term stable hydrogen sensors that can prevent major accidents.
Several different sensing platforms have been proposed, typically based on the change in optical \cite{LanZorKas07, LanLarKas10, LiuTanHen11, BoeBanSet17, NugDarZhd18, SheSheWan19, BenYamKur19, NugDarCus19, SteStrBot20, BanSchDam21, LosGutSuv21} or electrical \cite{HasIftChu16, KabChaTua16, AleSabKan20} properties of a material during hydrogen absorption.
Many of these devices are based on palladium (Pd), which forms a hydride phase upon exposure to a hydrogen-rich atmosphere.
The electronic properties of the hydrogenated system differ from the hydrogen-free one, leading to a shift in the optical response as well as a change in resistivity, which can be measured and related to the hydrogen pressure in the surrounding environment.
In this work, we focus on optical sensors based on the \gls{lspr} of Pd--Au-based nanoalloys \cite{LanZorKas07, LanLarKas10, NugDarZhd18, NugDarCus19} and show how they can be optimized via geometry and alloy composition.

A significant shortcoming of hydrogen sensors based on pure Pd \glspl{np} is that they undergo a first-order phase transition upon hydrogenation, from the hydrogen-poor $\alpha$ to the hydrogen-rich $\beta$-phase, which leads to a non-linear, discontinuous response to hydrogen pressure.
The phase transition is further associated with significant hysteresis \cite{GriStrGie16}, which ultimately makes the response of the sensor dependent on its history.
It has been shown that these disadvantages can be overcome by alloying \cite{WadNugLid15, DarNugLan21}.
Specifically, the introduction of about 20\% of gold (Au) suppresses the phase transition, making the response of the sensor a linear function of hydrogen pressure \cite{MaeFla65, WesRooLec13, WadNugLid15, NugDarZhd18, DarKhaTom21}.
To optimize these systems for hydrogen sensing, important questions remain: in what proportions should we mix the two alloyants, what shapes should the \glspl{np} have, and what features of the optical response should the sensing mechanism be based on?
Historically, similar questions have been addressed by trial-and-error experiments.
While that is a viable approach, the additional degrees of freedom introduced by mixing two or more components and the large space of available geometries make it extraordinarily cumbersome to experimentally test all combinations of potential interest.
In this context, a multi-scale modeling approach that resolves the variation in electronic properties with composition and translates this variation to the optical properties of \glspl{np} of different sizes and shapes is therefore invaluable.

The \gls{df} links electronic and optical properties, as well as atomic and continuum scales.
We have recently calculated the \glspl{df} of ten binary alloys in their full composition range and demonstrated that electrodynamic simulations based on these \glspl{df} provide accurate predictions of the optical response of alloy nanoparticles \cite{RahTibRos20}.
Earlier studies of pure Pd have shown, in principle, how one can combine electronic structure calculations and electrodynamic simulations to enable multi-scale modeling of the optical response during hydrogenation \cite{PoySilChe12, SilMuiChe12}.
They were, however, limited in the representation of chemical order (due to very small structure models) and also did not consider Pd-alloys.
Here, we build on and extend the underlying approach to calculate the \glspl{df} of Pd--Au alloys with up to 46\% Au and in the full range from no hydrogen to full hydrogen loading.
The \glspl{df} are then converted to a Lorentzian representation and used in \gls{fdtd} simulations to predict the extinction spectra of Pd-Au nanodisks.
Finally, the optical sensitivity of hydrogen sensors based on Pd--Au \glspl{np} is derived from the changes in their extinction spectra.
To obtain the actual sensitivity one must also include a thermodynamic factor that accounts for the solubility of hydrogen in the material, which here is taken into account using a thermodynamic model developed previously \cite{RahLofFra21}.
This multi-scale approach, which extends from first-principles calculations on the atomic scale to electrodynamic simulations on the macroscale, is not limited to hydrogenated Pd--Au nanoparticles, but can in principle be applied to the optical response of any material.

An important aspect in this context is atomic scale ordering, which has the potential to significantly alter the \gls{df} and, in turn, the optical response \cite{RahTibRos20}.
This situation requires careful consideration when calculating \glspl{df} from first principles and applying them to the prediction of the optical response of nanoalloys.
In thermodynamic equilibrium and in the absence of hydrogen, bulk Pd--Au is expected to form an alloy without long-range order and at most a low degree of short-range order \cite{LeeNohFla07, RahLofFra21}.
Furthermore, since in many studies the nanoalloys are annealed at high temperatures \cite{WadNugLid15, NugDarZhd18, DarKhaTom21}, it is safe to assume that the chemical distribution of the alloyants is close to random.
To represent this situation in our calculations, we use \glspl{sqs} \cite{ZunWeiFer90}, which are constructed to reproduce the (lack of) short-range order in a truly random alloy using unit cell sizes that are small enough for efficient calculations (in our case 24 Au/Pd atoms and 0--24 H atoms).
As discussed before \cite{LeeNohFla07, ChaSho14, RahLofFra21}, exposure to hydrogen can, however, impact thermodynamics and ordering.
As the nanoalloys are typically only exposed to hydrogen at high temperatures during fabrication and for short periods of time during operation, it is reasonable to assume that this effect is negligible for the samples produced in, \textit{e.g.}, Refs.~\citenum{WadNugLid15, NugDarZhd18, DarKhaTom21}.
It is, however, known from both experimental and theoretical evidence that an ordered, intermetallic phase can emerge for compositions around 25\% Au after annealing at high hydrogen pressures and moderate temperatures \cite{LeeNohFla07, RahLofFra21}.
As we will show in this work, the presence of this phase could potentially improve the sensing ability at low hydrogen pressures radically.

\section*{Results and Discussion}

\subsection{Electronic structure and dielectric functions}
\label{sect:dielectric-functions}

\begin{figure}
    \centering
    \includegraphics{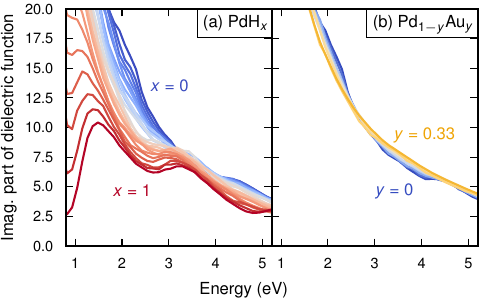}
    \includegraphics{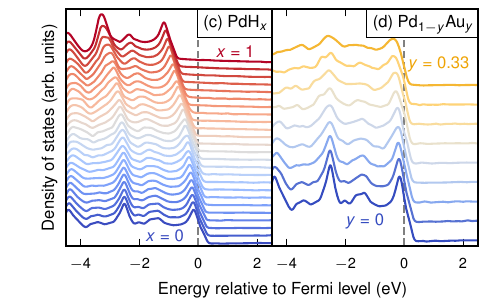}
    \caption{
    Variation of the imaginary part of the dielectric function as a function of photon energy for (a) PdH$_x$ without Au and (b) Pd$_{1-y}$Au$_y$ in absence of H, as calculated in this work.
    The dielectric function is closely related to the density of states (c--d).
    Energies have been shifted such that the Fermi level is always at \unit[0]{eV}.
    }
    \label{fig:dos}
\end{figure}

We begin by examining the \glspl{df} as calculated with \gls{lrtddft} (\autoref{fig:dos}a--b; see \autoref{sfig:df-all-real} and \autoref{sfig:df-all-imag} for the real and imaginary parts at more compositions and in a wider energy range).
A few trends are apparent.

In the case of Pd--H, addition of H leads to the emergence of two peaks in the imaginary part of the \gls{df} that at 100\% H loading are located around \unit[1.5]{eV} and at approximately \unit[3.0--3.5]{eV} (\autoref{fig:dos}a).
Comparison with the \gls{dos} indicates that these two features are related to two pronounced peaks in the d band (\autoref{fig:dos}c; see \autoref{sfig:dos-all} for the \gls{dos} at more compositions).
In fact in this case, the \gls{dos} in its entirety is almost unaffected by H addition, save for a rigid shift of the energy scale.
This is a result of the localized character of the H states, which do not hybridize with each other.
Addition of H also reduces the magnitude of the real part of the \gls{df} in the visible range, as has been previously observed in both experiment \cite{PalMurNar18, VarAzoCla14} and computations \cite{SilMuiChe12} (\autoref{sfig:df-experimental-comparison}).
Generally speaking, such a trend is expected to imply a redshift of the plasmon peak, as the Fröhlich condition for \gls{lspr} emergence is satisfied at lower energies.

Upon addition of Au to Pd, there is almost no change in the imaginary part of the \gls{df} in the composition range considered here (\autoref{fig:dos}b).
In the \gls{dos}, on the other hand, a notable smoothing occurs as a result of hybridization between the d states of Pd and Au (\autoref{fig:dos}d).\footnote{
    It is only at higher Au content that the distance between the d band and the Fermi level becomes significant \cite{NahJunKim98, RahTibRos20}.
}
Therefore, when H is added to Pd--Au, the two peaks that were apparent in the \gls{df} of Pd--H are largely smoothed out due to alloy disorder (\autoref{sfig:df-all-imag}).

\subsection{Optical response}
\label{sect:optical-response}

\begin{figure}
    \centering
    \includegraphics{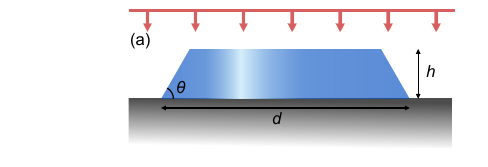}\\
    \includegraphics{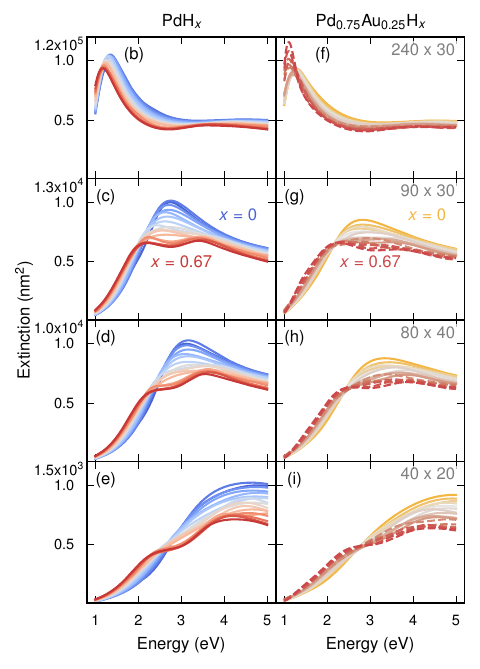}
    \caption{
        Optical response of truncated cone nanodisks of varying height $h$ and diameter $d$ and fixed cone angle $\theta=\unit[60]{^\circ}$ (a).
        Here, we show the extinction spectra for disk geometries ($d \times h$ in \unit{nm}): $240 \times 30$ (b, f), $90 \times 30$  (c, g), $80 \times 40$ (d, h) and $40 \times 20$ (e, i) with 0\% Au (b--e) and 25\% Au (f--i) at H concentrations 0--67\%.
        The dashed lines correspond to H concentrations $c_\text{H} > 0.67 - c_\text{Au}$, which are excluded in the calculations of the optical sensitivity $\widetilde{S}$.
    }
    \label{fig:AuPdH-spectra}
\end{figure}

After fitting the \glspl{df} to a Lorentzian representation (as described in \autoref{snote:lorentzian-rep}) we carry out \gls{fdtd} simulations of the optical response of truncated cone nanodisks (\autoref{fig:AuPdH-spectra}a), characterized by their height $h$ (20--\unit[40]{nm}) and \gls{ar} (2--12) or diameter $d=\text{AR}\times h$ on a silica substrate, mimicking a single nanodisk fabricated by the hole-mask colloidal lithography method \cite{FreAlaDmi07}.
The exact shape of the nanodisk is, however, not of great importance (see \autoref{sfig:fdtd-shape-effects} for a comparison between slightly differently shaped nanodisks).

\subsubsection{Pd--H nanodisks}

We begin by studying the optical response of the Pd--H system based on the calculated extinction.
In \autoref{fig:AuPdH-spectra}b--e the extinction spectra for four specific geometries are shown (see \autoref{sfig:AuPdH-spectra-abs-sca} for separate scattering and absorption spectra).
These are selected since they represent different characteristics of the studied spectra (see \autoref{sfig:extinction-heatmap} for the full range of geometries).
The shape of the extinction spectra varies with geometry and H content, but at least one peak can be identified for all systems.
Most geometries considered have extinction spectra similar to \autoref{fig:AuPdH-spectra}b with a well-defined, relatively sharp peak that shifts to the red with increasing H content.
For smaller nanodisks, however, the features are broader and we can identify two distinct peaks at H concentrations above 50\% (\autoref{fig:AuPdH-spectra}c--e).
The second peak emerges with increasing H content and decreasing height and diameter, and becomes the dominant feature in the high H loading limit for the smallest particles considered here.

\begin{figure}
    \centering
    \includegraphics{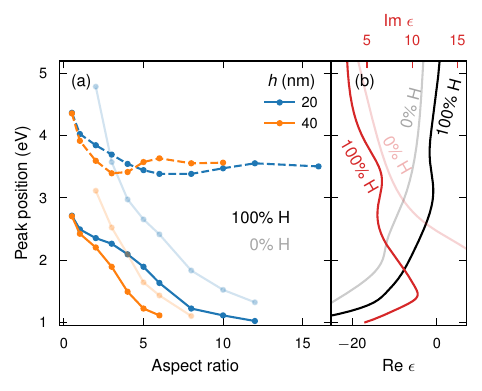}
    \caption{
        Peak position of the two peaks in the extinction spectra for PdH (opaque lines) and Pd (transparent lines) as a function of aspect ratio for two disk heights $h$ (a) and the corresponding dielectric functions (b).
        PdH displays two features; one low-energy peak (solid line) which shifts with aspect ratio and one high-energy peak (dashed lines) which only shifts for small aspect ratios, indicating avoided-crossing behavior due to coupling between the peaks.
        The high-energy peak can be traced to the feature at \unit[3.5]{eV} in the dielectric function (b) for 100\% H (opaque line).
        This feature is not present for 0\% H (transparent line), and as a result, the \gls{lspr} peak can shift over a much wider energy interval (transparent lines in b).
        Note that panel (b) shows the fitted Lorentzian representation of the dielectric functions.
    }
    \label{fig:peak-shif}
\end{figure}

The high-energy feature is in fact present for all geometries in the high H limit, but for larger nanostructures (such as the one in \autoref{fig:AuPdH-spectra}b) this feature is less noticeable due to the larger amplitude of the low-energy peak.
Of these two peaks, the high-energy one is roughly constant in energy (and amplitude, see \autoref{sfig:PdH-peak-shift-SI}) while the low-energy peak shifts when the aspect ratio is varied.
To illustrate this behavior, we compare the cases of 0\% and 100\% H in Pd, which allows one to identify the key features in the \glspl{df} more easily due to the lack of disorder on the H sublattice.
For 0\% H in Pd, we observe one peak which shifts from \unit[1]{eV} for large nanodisks to \unit[5]{eV} for small nanodisks (\autoref{fig:peak-shif}a), in agreement with the expected size dependence of a \gls{lspr}.
For 100\% H in Pd, on the other hand, we observe two peaks (see the corresponding extinction spectra and field enhancement in \autoref{sfig:PdH-near-fields}).
For nanodisks with larger \glspl{ar}, the peaks are separated in energy and the low-energy peak increases in energy with decreasing \gls{ar} (as expected for a \gls{lspr}) while the high-energy peak remains at about \unit[3.5]{eV} (\autoref{fig:peak-shif}a).
The lack of size dependence for the high-energy peak indicates that it is caused by a bulk phenomenon, such as a spectrally localized interband transition \cite{Pak11, PirPakMil14}.
When a \gls{lspr} peak moves closer to an interband transition, they can couple and exhibit avoided crossing \cite{Pak11, PirPakMil14}.
This is precisely the behavior we observe for small \glspl{ar} when the shift of the low-energy feature is restricted (compared to the 0\% H case) and the high-energy peak starts to increase in energy.
The presumed interband transition can be related to the previously discussed feature in the \gls{df} close to \unit[3.5]{eV} (\autoref{fig:peak-shif}b and \autoref{sect:dielectric-functions}).
For 0\% H in Pd there is no such feature, and as a result, there is no avoided crossing for 0\% H and the \gls{lspr} peak can shift over the entire energy interval without interference.

\subsubsection{Pd--Au--H nanodisks}

We now turn to the impact of Au on the extinction spectra.
For comparison with previous experimental studies \cite{NugDarZhd18, DarKhaTom21}, we focus on 25\% Au which provides a trade-off between suppressing the hysteresis and avoiding the decrease in H absorption with increased Au concentration (\autoref{sfig:hydrogen-content-at-pressure}).
The introduction of 25\% Au has a relatively small effect on the optical response (\autoref{fig:AuPdH-spectra}).
Some changes in the extinction spectra can, however, be identified.
For the most part, introducing Au leads to a broader extinction spectrum with slightly lower peak amplitude and a weaker tendency for peak splitting (\autoref{fig:AuPdH-spectra}g--i).
This is consistent with the observation that the \glspl{dos} and the \glspl{df} are typically smoothed by adding Au to Pd; features are blurred out by the chemical disorder.
In the case of large \gls{ar} and high H content, however, the \gls{lspr} peak becomes sharper and more intense with increasing Au content (\autoref{fig:AuPdH-spectra}f).
This sharpening is a consequence of the increased distance between d band and Fermi level in systems with high Au and/or H content (\autoref{fig:dos}c), which reduces the ability of the d electrons to screen the \gls{lspr} at low energies.
Peak splitting occurs in the high H limit for Pd--Au nanodisks as well, but the evolution of the high-energy peak with size or H content is considerably less regular compared to the case of Pd.
This is related to the corresponding somewhat irregular evolution of the alloy \glspl{df} which is caused by the imperfect representation of the chemical order due to limited system size.

\subsection{Optical sensitivity optimization}
\label{sect:sensitivity}

\begin{figure*}
    \centering
    \includegraphics{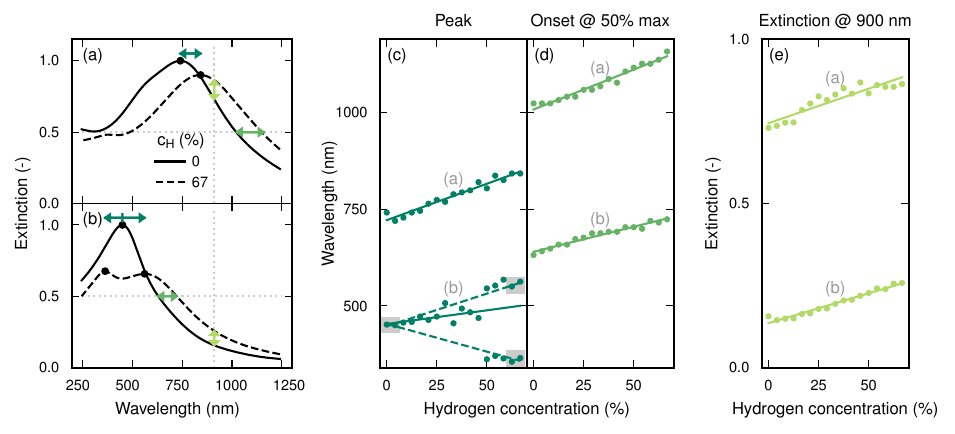}
    \caption{
        For a given set of extinction spectra, such as the spectra (a) without or (b) with pronounced peak splitting, one can define the optical sensitivity $\widetilde{S}$ based on several different shifts with H content (illustrated here for Pd--H).
        First, the peak position (black dots in (a,b)) generally shifts linearly with H concentration (c).
        For the case of pronounced double peaks, we study the peak shift both excluding data points with double peaks (solid lines in (c)) and with double peaks (dashed lines in (c)).
        In the latter case, only the data points corresponding to the two lowest and two highest H concentrations (indicated by the shaded regions) contribute to the fit.
        Second, the wavelength at a certain onset amplitude (50\% of the peak extinction at 0\% H here) have a similar linear increase with increased H content (d).
        Third, the extinction at a particular wavelength (\unit[900]{nm} here) also show an almost linear dependence on the H concentration (e), although the trend deviates slightly from linearity close to the peak(s).
        In all these cases, the optical sensitivity $\widetilde{S}$ can be defined as the slope of the linear trend(s).
    }
    \label{fig:sensitivity-principles}
\end{figure*}

In general, sensitivity is defined as the ratio between the output signal and the underlying external stimulus.
In \emph{optical} sensing it is common to follow the variation of one or more characteristic descriptors of the optical spectrum as measures of the output signal, which in the present case of plasmonic sensing are related to the behavior of the plasmon peak, \textit{e.g.}, its amplitude, width or position, or the surrounding spectral region.
In the specific case of hydrogen sensing the external stimulus is the presence of \ce{H2} in the surrounding environment as expressed by its partial pressure $p_{\mathrm{H}_2}$.
More formally the sensitivity of such a sensor, using here the position of the plasmon peak $\lambda_\text{max}$ for illustration, can be written as
\begin{align}
S = \frac{d\lambda_\text{max}}{dp_{\mathrm{H}_2}}.
\end{align}
Earlier measurements for this material system \cite{LanZorKas07, LanLarKas10, NugDarZhd18, NugDarCus19} have, however, commonly considered the change in the absorption maximum $\lambda_\text{max}$ with the hydrogen concentration in the material $c_\text{H}$,
\begin{align}
\widetilde{S} = \frac{d\lambda_\text{max}}{dc_{\mathrm{H}}}.
\label{eq:optical-sensitivity}
\end{align}
The two measures are related to each other via
\begin{align}
S = \frac{d\lambda_\text{max}}{dc_{\mathrm{H}}}
\times
\frac{dc_\text{H}}{dp_{\mathrm{H}_2}}
=
\widetilde{S} \, S_\text{sol}
\label{eq:sensitivity}
\end{align}
where the subscript emphasizes that $S_\text{sol}$ directly depends on the solubility of hydrogen as a function of its partial pressure.

In the present section, we will focus entirely on $\widetilde{S}$, for brevity referred to as the optical sensitivity.
We will return to the role of $S_\text{sol}$ in the \autoref{sect:thermodynamic-optimization}.

In the following, we study three specific measures for the optical sensitivity $\widetilde{S}$ based on the shift in peak position (\textit{peak shift}), wavelength at a fixed extinction amplitude onset (\textit{onset shift}), and extinction at a certain wavelength (\textit{extinction shift}), respectively (\autoref{fig:sensitivity-principles}), with respect to the absorbed H concentration to determine the limits of optical sensitivity optimization through engineering of composition and/or geometry.

\begin{figure}
    \centering
    \includegraphics{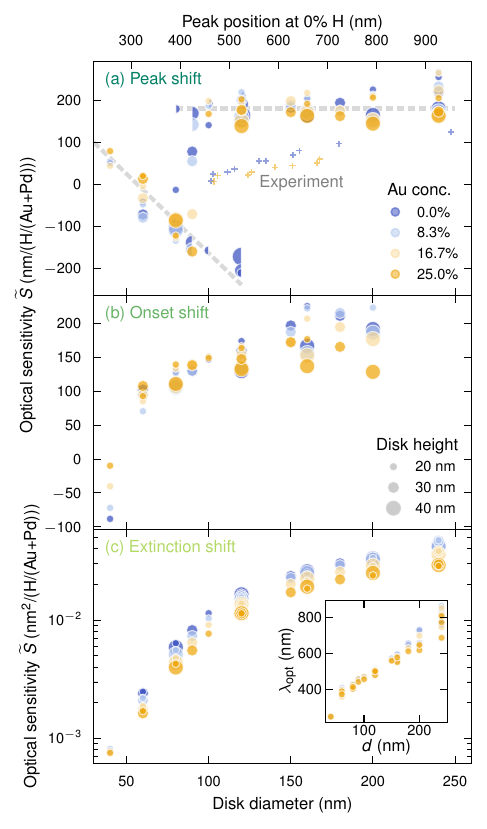}
    \caption{
        Optical sensitivity of Pd--Au nanodisks with varying Au content as a function of the disk diameter, in terms of (a) peak shift, (b) onset shift and (c) extinction shift at $\lambda_{\text{opt}}$ per unit H uptake.
        In (a), semi-circles corresponds to the sensitivity when peak-splitting is taken into account (as explained in \autoref{fig:sensitivity-principles}) and plus signs show experimental data from Ref.~\citenum{NugDarZhd18}.
        Note that the experimental sensitivity is rescaled from nm/(H/Pd)) to nm/(H/(Au+Pd)) and the disk diameter is approximated by the linear relation in \autoref{sfig:sensitivity-lambda0}b.
        The corresponding peak position at 0\% H is indicated by the upper x-axis scale, calculated from the diameter and the linear relation in \autoref{sfig:sensitivity-lambda0}b.
        In (c), the inset shows the wavelength where the optimal sensitivity was found for each system.
    }
    \label{fig:PdAuH-sensitivity}
\end{figure}

In the case of Pd--H, the H concentration range that is relevant under common thermodynamic conditions is rather wide, while it is more narrow for Pd--Au--H.
As a consequence, we have chosen to restrict the maximum H concentration considered to $c_\text{H} < 0.67 - c_\text{Au}$ when calculating the sensitivities, which is sufficient for hydrogen pressures up to \unit[1]{bar} (\autoref{sfig:hydrogen-content-at-pressure}).
The exact limit is not of great importance since the studied shifts are generally linear with H concentration for hydrogen pressures $\lesssim \unit[1]{bar}$.

\subsubsection{Peak shift}

In Ref.~\citenum{NugDarZhd18}, the (optical) sensitivity was defined as the ratio of the peak shift to the change in H concentration.
For a specific geometry, the peak position is generally an approximately linear function of the H concentration (\autoref{fig:sensitivity-principles}c), and the sensitivity can be obtained as the slope of the corresponding linear fit.
This is straightforward for the larger nanostructures considered (\textit{i.e.} disks well-represented by \autoref{fig:sensitivity-principles}a) that feature one dominant peak for all H concentrations.\footnote{In the sensitivity analysis, we exclude all peaks below the mean extinction over the spectrum since those are in practice not of importance.}
For the smaller nanodisks considered here, peak splitting, however, comes into play which complicates the determination of the optical sensitivity.
In systems with significant peak splitting (see, \textit{e.g.}, \autoref{fig:sensitivity-principles}b), we therefore consider two different peak shift sensitivities.
First, we fit the peak shift as a function of H concentration after excluding all H concentrations with peak splitting (solid line in \autoref{fig:sensitivity-principles}c).
Note that with this approach, cases with pronounced peak splitting typically result in a large fitting error since the data points are characterized by large scatter (see \autoref{sfig:PdH-sensitivity-rmse} for the fitting errors).
Second, we fit the peak shift based on only the first two data points (low H limit) and the last two data points (high H limit) for either peak separately (dashed lines in \autoref{fig:sensitivity-principles}c), resulting in two sets of sensitivities, typically with opposite signs.
The latter approach is suitable for pure Pd and low Au content since the phase transition introduces a jump from low to high hydrogen content (\autoref{sfig:hydrogen-content-at-pressure}).
For high Au content ($\gtrsim\!10\%$), peak splitting is in practice not an issue since the hydrogen content is below 50\% at relevant pressures (\autoref{sfig:hydrogen-content-at-pressure}).

In \autoref{fig:PdAuH-sensitivity}a, we display the obtained peak shift sensitivity for all considered systems as a function of the disk diameter.
Clearly, the size dependence of the sensitivity can be divided into two regimes.
Nanodisks with diameter $\gtrsim\!\unit[100]{nm}$ have a roughly constant sensitivity around \unit[180]{nm/(H/(Au+Pd))}, which to a large degree is independent of disk height and alloy composition.
Smaller nanodisks, however, follow a linearly decreasing trend with increasing diameter.
The double peak sensitivities (semi-circles in \autoref{fig:PdAuH-sensitivity}a) follow the same trends and introduce an overlap between the two regimes.
The trend for small nanodisks can be explained by the peak splitting.
When the high-energy peak at 67\% H ($\lambda_{\text{H}_\text{max}}$) dominates over the corresponding low-energy peak, the sensitivity is determined by the shift between the 0\% H peak ($\lambda_{0}$) and the former.
In a simplified picture, $\lambda_{0}$ shifts to lower wavelengths with decreasing diameter while $\lambda_{\text{H}_\text{max}}$ is constant.
This means that, as the diameter is decreased, the sensitivity will shift from strongly negative to positive, crossing zero when $\lambda_{0}$ and $\lambda_{\text{H}_\text{max}}$ align.

The optical sensitivity trends are largely unaffected by the introduction of up to 25\% Au (\autoref{fig:PdAuH-sensitivity}a).
This is somewhat surprising since for Pd the sensitivity is almost entirely determined by peak splitting and for Pd--Au the peak splitting is less distinct and often not relevant due to the H concentration threshold.
Clearly, the  peak shift sensitivity is not very sensitive to the details in the shape of the extinction spectra.

In \autoref{fig:PdAuH-sensitivity}a, we include experimental results from Ref.~\citenum{NugDarZhd18}, where the authors find a linear increase in the (optical) sensitivity as a function of the peak position at 0\% H.\footnote{
    Here, the experimental peak position at 0\% H has been converted to disk diameter based on our obtained linear relationship (\autoref{sfig:sensitivity-lambda0}b) and the sensitivity is rescaled from H concentration with respect to Pd to the total H concentration.
    An equivalent comparison based on peak position at 0\% H can be found in \autoref{sfig:sensitivity-lambda0}a.
}
While the experimental trend can be interpreted to reach a plateau for peak positions at 0\% H $\gtrsim\!\unit[600]{nm}$, there is still an apparently large discrepancy with our findings.
Note, however, that the experimental results were obtained for nanodisk \emph{arrays} and the measurements thus correspond to an average over a large \emph{ensemble} of particles.
Notably, the measured (optical) sensitivity is very close to zero right at the discontinuity between the two sensitivity regimes, where peak-splitting is most pronounced.
This could suggest that the experimentally measured linear increase in sensitivity is due to a superposition of nanoparticle shapes and sizes that fall on either side of the discontinuity between the sensitivity regimes, and as the fraction of particles which display peak splitting decreases, the sensitivity increase.
We address this aspect in detail in \autoref{sect:broadening} but find that size distribution and array effects on their own are insufficient to explain this effect.

\subsubsection{Onset shift}

Another way of defining the optical sensitivity, which circumvents keeping track of peaks, is to monitor the wavelength at the onset of a fixed target extinction amplitude.
Here, we define the target amplitude as 50\% of the peak amplitude at 0\% H (\autoref{fig:sensitivity-principles}a,b) and track the corresponding wavelength (to the right of the peak in wavelength) as a function of the H concentration (\autoref{fig:sensitivity-principles}d).\footnote{
    The definition of the onset is arbitrary and does not have to be based on the peak. Here we use 50\% of the peak amplitude as a convenient measure when comparing sensitivity over the span of geometries.
}

The onset shift generally follows a linear trend with respect to H concentration, and the slope of which can be defined as the sensitivity (\autoref{fig:PdAuH-sensitivity}b).
For almost all considered geometries, such as the ones in \autoref{fig:sensitivity-principles}a--b, we obtain a positive sensitivity that increases slightly with increasing diameter.
For very small nanostructures, however, the low energy peak at high H content falls below the onset amplitude  (\autoref{fig:AuPdH-spectra}e,i), which reverses the direction of the onset shift and results in a small or negative sensitivity.
In principle, this issue could be resolved by decreasing the onset amplitude value below 50\%.
That would, however, push the amplitude onset for the largest nanodisks beyond the low energy (high wavelength) limit.

The introduction of Au slightly reduces the slope of the sensitivity trend such that the sensitivity of the largest nanodisks decreases with increased Au content.

\subsubsection{Extinction shift at a certain wavelength}

A third possible optical sensitivity definition is based on the shift in extinction amplitude at a certain wavelength (\autoref{fig:sensitivity-principles}a, b, e).
This definition is of particular interest for real devices since it does not require measuring the spectrum over the entire spectral range but only at a single wavelength.
In \autoref{sfig:extinction-shift}, we display extinction spectra at increasing H content in terms of extinction at selected energies (\autoref{sfig:extinction-shift}a--d) and the corresponding slope of linear fits (\autoref{sfig:extinction-shift}e--h).

The slope changes significantly with energy from a positive value to the left of the peak to a negative value to the right of the peak.
This means that the choice of wavelength is crucial for maximizing the optical signal.
For each system we define the wavelength for which the largest slope (in absolute value) is obtained as $\lambda_\text{opt}$.
The extinction shift sensitivity is defined as the absolute value of the slope at $\lambda_\text{opt}$ in units corresponding to extinction cross section (\autoref{fig:PdAuH-sensitivity}c).
The sensitivity increases with disk diameter, decreases slightly with Au content and, in the size range considered her, is independent of disk height.
It should be noted that, in contrast to peak and onset shift, the extinction shift is proportional to the amplitude of the optical response of an individual nanodisk.
As a result, the sensitivity increase is to a large degree caused by the increase in extinction cross section with size (\autoref{fig:AuPdH-spectra}).
In a real device, the number of nanodisks per area can be increased for smaller particles which means that for some purposes, it might be more relevant to compare the extinction-shift sensitivity scaled with nanodisk area.

\subsubsection{Uncertainty analysis}\label{sect:broadening}

As previously noted, there is a discrepancy between our results and an earlier experimental study \cite{NugDarZhd18} for the peak shift sensitivity.
Such differences between experiment and simulations are not unexpected given the many uncertainties involved, which exist both at the level of the electrodynamic simulations of the optical response and at the level of the electronic structure calculations of the \glspl{df}.
Here, we investigate the impact of the sources of discrepancy that we judge to be the most important.

\begin{figure}
    \centering
    \includegraphics{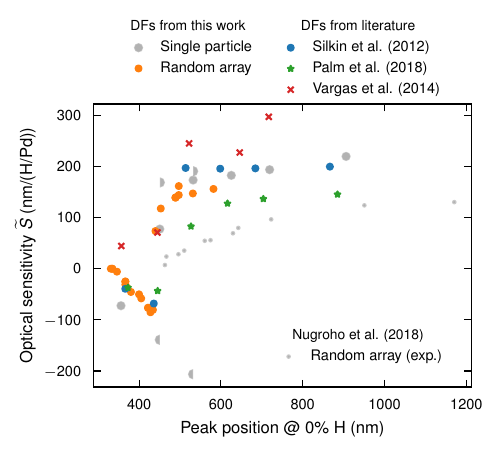}
    \caption{
        Optical peak shift sensitivity $\widetilde{S}$ for Pd nanodisks with \unit[30]{nm} diameter calculated using different  computational setups and \glspl{df}.
        The data from \autoref{fig:PdAuH-sensitivity}a is included in gray.
        The sensitivities for explicit and effective random arrays (orange), with and without size distribution, follows closely the previously discussed trends for single particles (large gray (semi-)circles).
        The sensitivities obtained using \glspl{df} from Silkin \textit{et al.} \cite{SilMuiChe12}, Palm \textit{et al.} \cite{PalMurNar18} and Vargas \textit{et al.} \cite{VarAzoCla14} show a large spread.
        Note that, except for the data in gray, the sensitivities here are calculated based on only the peaks at 0\% and $\sim\!67\%$ H \footnote{The exact concentration of the hydride is 67\% for the Silkin \textit{et al}. \gls{df}, 62\% for the Palm \textit{et al.} \gls{df}, and 58 \% for the Vargas \textit{et al.} \gls{df} due to the concentrations of the available data.} in order to reduce the computational effort.
    }
    \label{fig:sensitivity-comparison}
\end{figure}

Let us first consider the effect of particle size distribution and array effects on the optical spectrum, which enter at the level of the electrodynamic simulations.
In contrast to the single particles simulated in this work, the experimental samples consist of many irregularly shaped nanoparticles with a significant size distribution \cite{NugDarZhd18}.
These particles are furthermore randomly distributed on a substrate, leading to array effects that are known to influence the optical spectrum \cite{AntTar15}.
We observe that the naive approach of weighting extinction spectra according to some size distribution does not significantly alter the extinction (\autoref{sfig:size-distribution}).
To better mimic the experimental situation we therefore performed additional simulations of random arrays of nanoparticles (using the T-matrix method as described in \autoref{sect:tmat}).
While the array effects (but to a lesser extent the size distribution) do change the position and sharpness of the peak (\autoref{sfig:array_res}; also see Ref.~\citenum{Czajkowski:20}), the sensitivity trends remain similar to the single particle results (\autoref{fig:sensitivity-comparison}).

A further source of uncertainty are the \glspl{df}, which in the present work are obtained from electronic structure calculations.
The calculated \gls{df} depends on the underlying electronic structure method, in the present case primarily the exchange-correlation functional, but also on the treatment of scattering, especially due to defects (including, \textit{e.g.}, surfaces, grain boundaries, and dislocations) and chemical disorder in the case of alloys.
With regard to the description of the electronic band structure, the present approach yields results that are in very good agreement with recent highly accurate many-body theory calculations \cite{VilLeiMar22}.
We therefore focus our attention on the possible effect of scattering mechanisms.
In \autoref{fig:sensitivity-comparison}, we compare the optical sensitivity calculated using three sets of \glspl{df} from the literature; two experimentally measured ones (Vargas \textit{et al.}\cite{VarAzoCla14} and Palm \textit{et al.} \cite{PalMurNar18}) and one calculated using an approach similar to ours (Silkin \textit{et al.} \cite{SilMuiChe12}).
Although the \glspl{df} are in semi-quantitative agreement over the energy range considered (\autoref{sfig:df-experimental-comparison}), small differences clearly can have a large effect on the resulting sensitivity.
Note in particular the large span of sensitivities resulting from the \emph{experimentally measured} sets of \glspl{df}, while the calculated set of \glspl{df} yield sensitivities very similar to our original findings.
Real samples are typically subject to defect scattering which leads to a broadening of the Drude peak.
This effect is very difficult to include rigorously at the ab-initio level and is furthermore sample-dependent, which could explain the large difference between the two experimentally measured sets of \glspl{df}.
To illustrate this aspect further, it is instructive to consider the case of the noble metals, for which the d-band feature in the \glspl{df} is clearly separated from the Drude peak and thus allows one to observe the effect of defect scattering more directly.
This reveals the large variation in the width of the Drude peak caused by differences in sample preparation and the resulting defect density (see, \textit{e.g.}, Figure~S1a--c of Ref.~\citenum{RahTibRos20}).
In the case of Pd alloys and hydrides the d-band overlaps with the Fermi level, which leads to a very broad response in the low-energy region and the impact of (defect) scattering cannot be clearly separated.
This makes it difficult to include this contribution systematically at the level of the \glspl{df} and eventually assess its impact on the optical response.

Regardless of which set of \glspl{df} is used, a large shift in sensitivity appears when the peak position at 0\% H is at approximately \unit[450]{nm}, which is also where the experimentally measured sensitivity goes to zero, in line with the discontinuity between the two sensitivity regimes previously discussed.
This indicates that it is in fact the interband transition at high H content that is responsible for large changes of the sensitivity.
We verified this argument by explicitly removing the associated Lorentzian from the \gls{df}, which results in a sensitivity that remains at a high, positive value even as the peak position at 0\% moves below \unit[450]{nm} (\autoref{sfig:remove-IBT}).

Lastly, we observe that by introducing artificial Lorentzian broadening in the simulated extinction spectra, a linear increase in the sensitivity similar to the experimental results can be obtained (\autoref{sfig:PdH-broadening}), although quite extreme levels of broadening are necessary.
We conclude that it is in general very difficult to account quantitatively for the sample-specific broadening channels.

\subsection{Thermodynamic optimization}
\label{sect:thermodynamic-optimization}

Above, in agreement with earlier experimental studies, we have seen that the alloy composition is not crucial for the \emph{optical} sensitivity $\widetilde{S}$.
It is, however, crucial for the \emph{thermodynamic} sensitivity $S_\text{sol}$ and hence impacts the actual sensitivity $S$, see Eq.~\eqref{eq:sensitivity}.
In this section, we therefore analyze the dependence of the \emph{thermodynamic} sensitivity $S_\text{sol}$ on alloy composition and then translate the results to a detection limit in terms of the hydrogen pressure.
In this context, we also show that the highest sensitivities are obtained for the ordered L1$_2$-\ce{Pd3Au} phase.

\subsubsection{Thermodynamic sensitivity}

\begin{figure}
    \centering
    \includegraphics{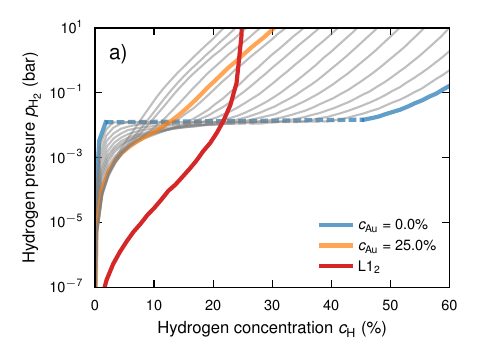}
    \includegraphics{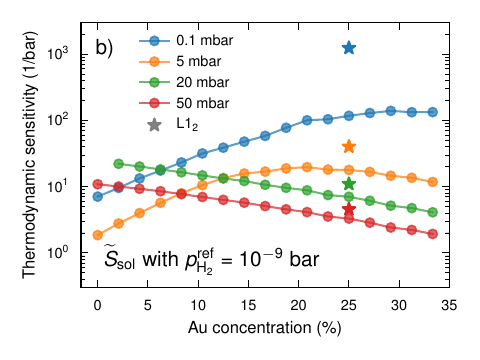}
    \caption{
        (a) Hydrogen partial pressure as a function of hydrogen concentration at a temperature of \unit[300]{K} for different gold concentrations according to Ref.~\citenum{RahLofFra21}.
        The dashed lines indicate the miscibility gap between the H-poor $\alpha$-phase and the H-rich $\beta$-phase.
        (b) Thermodynamic sensitivity $\widetilde{S}_\text{sol}$ according to Eq.~\eqref{eq:thermodynamic-sensitivity} using a reference pressure of $p_{\mathrm{H}_2}^\text{ref}=\unit[10^{-9}]{bar}$ as a function of Au concentration at different \ce{H2} partial pressures and a temperature of \unit[300]{K}.
    }
    \label{fig:thermodynamic-sensitivity}
\end{figure}

For the following analysis, we limit us to a temperature of \unit[300]{K} and use the relationships between partial pressure of \ce{H2}, Au concentration, and H concentration established in Ref.~\citenum{RahLofFra21} by means of alloy cluster expansions, Monte Carlo simulations, and experimental data (\autoref{fig:thermodynamic-sensitivity}a and \autoref{sfig:hydrogen-content-at-pressure}).
Using these data one can compute the (differential) thermodynamic sensitivity (also see Eq.~\eqref{eq:sensitivity})
\begin{align}
    S_\text{sol} = \frac{dc_\text{H}}{dp_{\mathrm{H}_2}}.
    \label{eq:differential-thermodynamic-sensitivity}
\end{align}
In practice one is, however, more often interested in the change relative to a reference pressure, representing a situation in which one aims to probe a sudden increase in hydrogen pressure relative to a low background level.
Here, we therefore consider the thermodynamic sensitivity defined as
\begin{align}
    \widetilde{S}_\text{sol}
    = \frac{
        c_\text{H}(p_{\mathrm{H}_2}) - c_\text{H}(p_{\mathrm{H}_2}^\text{ref})
      }{
        p_{\mathrm{H}_2} - p_{\mathrm{H}_2}^\text{ref}
      }.
    \label{eq:thermodynamic-sensitivity}
\end{align}
While there are some quantitative differences between these two measures, they show qualitatively the same behavior (\autoref{fig:thermodynamic-sensitivity} and \autoref{sfig:thermodynamic-sensitivity}).

The thermodynamic sensitivity varies with both Au concentration and \ce{H2} partial pressure (\autoref{fig:thermodynamic-sensitivity}b).
While at smaller partial pressures ($\lesssim\,\unit[1]{mbar}$) one obtains a monotonic \emph{increase} of the thermodynamic sensitivity with Au content, at larger partial pressures ($\gtrsim\,\unit[20]{mbar}$) one actually observes a \emph{decrease} of $\widetilde{S}_\text{sol}$ with Au content.
This behavior can be readily rationalized in terms of the pressure-concentration isotherms (\autoref{fig:thermodynamic-sensitivity}a).
In the intermediate region around approximately \unit[10]{mbar}, one observes a maximum, which shifts with hydrogen partial pressure.

Since the actual sensitivity $S$ is simply the product of the optical sensitivity $\widetilde{S}$ and the thermodynamic sensitivity $S_\text{sol}$, see Eq.~\eqref{eq:sensitivity}, and the optical sensitivity does not strongly depend on Au concentration, the behavior described above should dominate the composition dependence of the actual sensitivity.
One should, however, also take into account the fact that experiments are commonly carried out for ensembles of particles, for which the $p_{\text{H}_2}$--$c_\text{H}$ isotherms are considerably broadened.
As a result, the variation of both the thermodynamic sensitivity ($S_\text{sol}$ or $\widetilde{S}_\text{sol}$) and the actual sensitivity ($S$) with Au concentration should be less pronounced than in the present case, which models the thermodynamic behavior of single particles under idealized conditions.

\subsubsection{Detection limit}

Next, we translate our results to a detection limit in terms of hydrogen pressure.
We assume a fixed optical sensitivity $\widetilde{S}$ as defined by the gray, dashed line in \autoref{fig:PdAuH-sensitivity}a (\unit[180]{nm/(H/(Pd+Au))}).
We then explore the hydrogen pressure required to induce a certain \gls{lspr} peak shift (\autoref{fig:detection-limit}a).
In stable, experimental setups, a peak shift of \unit[0.1]{nm} or less can be detected fairly easily.
This implies that the detection limit is well below \unit[1]{mbar} (green line in \autoref{fig:detection-limit}a), which is approximately the target (\unit[1,000]{ppm}) set by the US \gls{doe} for hydrogen sensors \cite{DoE15} (dashed, gray line in \autoref{fig:detection-limit}a).
This is true regardless of Au content in the concentration interval studied here.
Under more challenging circumstances, however, it cannot be ruled out that a more significant peak shift is required to obtain a reliable signal.
With a \unit[1]{nm} peak shift (orange line), nanodisks of pure Pd do not meet the US \gls{doe} target, and at \unit[10]{nm} (blue line), the target cannot be met at all.

\subsubsection{Enhancing sensitivity through chemical ordering}

The situation can be substantially improved if we consider the ordered L1$_2$-\ce{Pd3Au} phase, which is expected to form when the alloy is subjected to high pressures of \ce{H2} \cite{RahLofFra21}.
This compound absorbs large amounts of hydrogen already at pressures below $\unit[10^{-5}]{bar}$, for which the regular alloy shows almost no sorption (\autoref{fig:thermodynamic-sensitivity}a), as observed both in experiments \cite{LeeNohFla07} and calculations \cite{RahLofFra21} (also see \autoref{sfig:hydrogen-content-at-pressure}).
As a result, one obtains a substantially larger thermodynamic sensitivity than for the regular alloy, especially for pressures $\lesssim\,\unit[1]{mbar}$ (stars in \autoref{fig:thermodynamic-sensitivity}b).

\begin{figure}
    \centering
    \includegraphics{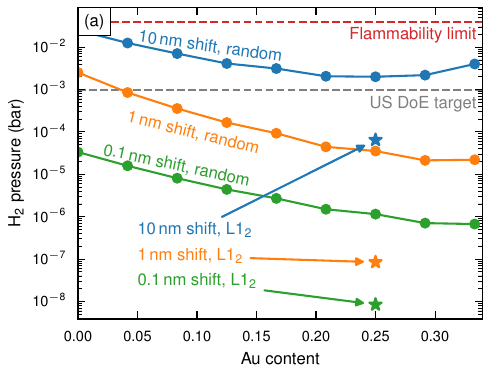}
    \includegraphics{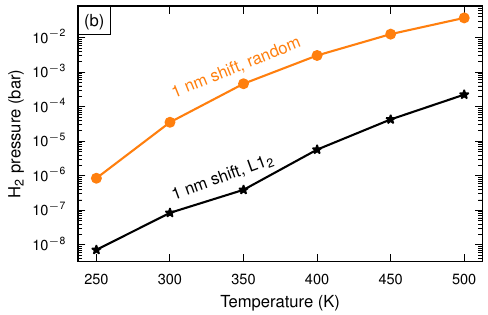}
    \caption{
        (a)~hydrogen pressure required to induce a peak shift of \unit[10]{nm} (blue), \unit[1]{nm} (orange), and \unit[0.1]{nm} (green) at \unit[300]{K} assuming a randomly ordered Pd--Au sublattice (dots) or L1$_2$ ordering (stars).
        Peaks shifts are calculated from the dashed, gray line in \autoref{fig:PdAuH-sensitivity}a (\unit[180]{nm/(H/(Au+Pd))}) based on the relation between \ce{H2} partial pressure and Au/H concentration calculated in Ref.~\citenum{RahLofFra21}.
        The flammability limit and the lower target for the detection limit of hydrogen sensors set by the US DoE \cite{DoE15} are marked in dashed red and gray, respectively.
        (b)~Temperature dependence for the detection limit assuming a peak shift of \unit[1]{nm} can be resolved for random Pd$_{0.75}$Au$_{0.25}$ (orange) and the same composition but with L1$_2$ ordering (black).
    }
    \label{fig:detection-limit}
\end{figure}

To evaluate the suitability of this L1$_2$-\ce{Pd3Au} compound for optical hydrogen detection, we calculated the dielectric function at different hydrogen contents and performed \gls{fdtd} simulations in the same fashion as for the disordered phase (\autoref{sfig:L12-spectra}).
The results are very similar for the ordered and the disordered phase, with almost identically positioned \gls{lspr} peaks.
We can thus assume the same sensitivity of with respect to the peak position as for the disordered phase.
The larger variation of the hydrogen uptake with pressure translates, however, to a detection limit that is predicted to be approximately one to two orders of magnitude lower in the ordered compound compared to the disordered phase, regardless of whether a large or small peak shift can be detected (stars in \autoref{fig:detection-limit}a).

When temperature is increased, a higher hydrogen pressure is required to maintain the same hydrogen content in Pd--Au.
Consequently, a higher hydrogen pressure is required to achieve a certain peak shift.
The detection limit thus decreases with increasing temperature.
The hydrogen uptake is, however, significantly higher in the L1$_2$-ordered phase than in the random phase, also at higher temperatures (\autoref{fig:detection-limit}b).
Hydrogen sensors based on the L1$_2$-ordered phase may thus retain a satisfactory detection limit at temperatures at which random alloys no longer meet the requirements.

\section*{Conclusions}

Our results indicate that for all optical sensitivity measures, the disk diameter is the most relevant descriptor (\autoref{fig:PdAuH-sensitivity}) in the selected span of geometries while disk height, \gls{ar}, and alloy composition are of secondary importance.
The highest optical sensitivities are found for larger particles, but for disk diameters above \unit[150]{nm} the improvement is limited.
For the optical sensitivity based on peak shift, our results reveal two separate regimes with respect to particle size (\autoref{fig:PdAuH-sensitivity}a).
For larger nanodisks ($d\gtrsim\unit[100]{nm}$), the optical sensitivity is approximately constant at \unit[180]{nm/$c_\text{H}$}.
For smaller nanodisks ($d\lesssim\unit[100]{nm}$), on the other hand, the optical sensitivity starts at a negative value and increases with decreasing particle size.
This behavior originates from the emergence of a second peak that appears for H concentrations above 50\%, caused by a localized interband transition, which becomes more apparent as the particle dimensions shrink.

% Block 3: Comparison to experiments
The peak shift sensitivities obtained from electrodynamic simulations using the calculated \glspl{df} are comparable in magnitude to direct experimental measurements.
They are, however, consistently larger and more importantly in the region where direct comparison is possible remain constant, whereas the experimental data shows a linear shift with the absorption maximum.
According to our calculations, while array effects and uncertainties in the particle size and shape distribution improve the agreement with experiment to some extent, the effect is too small to account for the deviation.
To assess the contribution of the \glspl{df}, we also calculated the optical sensitivities using \glspl{df} from experiment.
The results obtained using different experimentally measured \glspl{df} exhibit a spread that is comparable to the deviation between the optical sensitivities obtained using the calculated \glspl{df} and the direct measurements.
This suggests that additional contributions to the \glspl{df} associated with scattering by, \textit{e.g.}, defects are crucial.
This highlights that the calculated \glspl{df} are in principle suitable for predictive simulations but that more work is required to assess more quantitatively the impact of sample-specific scattering and broadening channels.

Regardless, our simulations thus suggest that the optical sensitivity of individual nanoparticles cannot be significantly improved beyond the plateau that has been demonstrated here.
For an array of nanoparticles, the optical sensitivity can, however, be improved by arranging them on a lattice and/or shifting the \gls{lspr} farther away from the avoided crossing, \textit{i.e.}, by increasing the nanodisk diameter far beyond \unit[100]{nm}.
In fact, the experimentally measured optical sensitivity in the far-infrared (\textit{i.e.}, for nanodisks with the largest diameters) clearly approaches the plateau value predicted here.
Importantly, while from the experimental data one might anticipate the optical sensitivity to increase further for an even more extreme shift to the infrared (regardless of whether it is experimentally realizable or not), the present simulations provide strong evidence that there is an upper bound that is intrinsic for this kind of material and approach.

The actual sensitivity $S$ comprises both the optical sensitivity $\widetilde{S}$ and a thermodynamic factor $S_\text{sol}$, which depends on the solubility of hydrogen as a function of partial pressure.
This thermodynamic sensitivity can vary rather strongly not only with the partial pressure of \ce{H2} but the Au concentration.
In particular for partial pressures near the $\alpha$/$\beta$ phase transition one can observe a strongly non-monotonic variation with Au concentration with a maximum near the Au concentration at which the $\alpha$/$\beta$ two-phase region closes.
While in experiments ensemble effects are likely to broaden this maximum, it remains a possible means to increase the actual sensitivity by alloying.

% Block 4: Thermodynamic punchline
Another strategy to improve a H sensor is to increase the H uptake at a fixed hydrogen pressure, which means that for a specific optical sensitivity (measured optical shift per unit of absorbed H), lower hydrogen pressures can be detected.
Here, we have shown that this can be achieved \textit{via} an ordered L1$_2$-\ce{Pd3Au} phase, which can be obtained through annealing at high hydrogen pressures.
Our simulations show that the optical response is largely unaffected by the presence of the ordered phase, which means that the optical sensitivity is unchanged.
The H uptake, on the other hand, is significantly higher for the ordered phase which yields a detection limit one to two orders of magnitudes lower than for the disordered phase, well below the threshold stipulated by the US Department of Energy (\autoref{fig:detection-limit}).
The possibilities provided by tuning the thermodynamic sensitivity through alloying demonstrated here have potentially notable implications for the field of plasmonic H sensing.

%%%%%%%%%%%%%%%%%%%%%%%%%%%%%%%%%%%%%%%%%%%
\section*{Methods}
\label{sect:methods}

\subsection*{Calculation of dielectric functions}\label{sect:tddft}
\Glspl{sqs} \cite{ZunWeiFer90} were generated with the \textsc{icet} package \cite{AngMunRah19} using the method and parameters suggested in Ref.~\citenum{WalTiwJon13}.
The ionic positions and cell shapes of these structures were relaxed with \gls{dft} in the projector augmented wave formalism as implemented in the Vienna \textit{ab initio} simulation package (version 5.4.1, PAW 2015) \cite{KreFur96b, KreJou99} using the vdW-DF-cx exchange-correlation functional \cite{BerHyl14}, until residual forces were below \unit[10]{meV/\AA} and stresses below \unit[1]{kbar}.
In these calculations, the wave functions were expanded in a plane wave basis set with a cutoff of \unit[384]{eV}, and the \gls{bz} was sampled with a $\Gamma$-centered grid with a $\boldsymbol{k}$-point density corresponding to $19\times19\times19$ $\boldsymbol{k}$-points in the primitive cell.
Occupations were set using the first-order Methfessel-Paxton scheme with a smearing parameter of \unit[0.1]{eV}.

\Glspl{df} were calculated using \gls{rpa}--\gls{lrtddft} as implemented in the GPAW package \cite{EnkRosMor10, YanMorJac11} (version 19.8.1 with GLLB-sc patched for extended metallic systems).
The macroscopic \gls{df} was calculated in reciprocal space through the linear density--density response function with wave functions expanded in a plane-wave basis with a cutoff at \unit[350]{eV}.
The \gls{bz} was sampled with a $\Gamma$-centered grid with a density corresponding to 61 $\boldsymbol{k}$-points in each direction for the undecorated (mono-elemental) primitive cell, and occupation numbers were smeared using a Fermi-Dirac distribution with a width of \unit[0.1]{eV}.

\Glspl{df} were obtained in the optical limit by evaluation at $\boldsymbol{q}=\boldsymbol{0}$.
As discrete $\boldsymbol{k}$-point sampling precludes intraband transitions with $\boldsymbol{q} \rightarrow \boldsymbol{0}$ and $\omega \rightarrow 0$, an intraband term, $\omega_\mathrm{P}^2 / \left(\omega + i \eta\right)^2$, was added to the \glspl{df}, where $\omega_\mathrm{P}$ is the calculated plasma frequency, using the broadening parameter $\eta=\unit[0.01]{eV}$.
The ground state wave functions were obtained with the GLLB-sc exchange-correlation functional \cite{KuiOjaEnk10}, whereas dynamic exchange-correlation effects were neglected, \textit{i.e.}, the \gls{rpa} approximation was used.
In the optical limit, \gls{alda} results approach the \gls{rpa} results, meaning virtually identical results would have been obtained if dynamic exchange-correlation effects had been taken into account using \gls{alda}.

\subsection*{FDTD simulations}\label{sect:fdtd}

Single particle extinction spectra were calculated using \gls{fdtd} simulations as implemented in the \textsc{meep} software \cite{OskRouIba10}.
The computational cell consists of a \ce{Pd_{1-y}Au_yH_x} nanodisk (represented by the calculated \glspl{df} fitted to a Lorentzian representation as described in \autoref{snote:lorentzian-rep}; Refs.~\citenum{EriFraErh19, Sch78, VirGomOli20}) placed on a \unit[100]{nm} thick SiO$_2$ substrate with constant refractive index 1.478.
The nanodisk has truncated-cone geometry with fixed cone angle $\theta=\unit[60]{^\circ}$, varying (bottom) diameter $d$ and height $h$ (\autoref{fig:AuPdH-spectra}a).
To account for the expansion that occurs with H absorption in a real system, the system size is scaled with H content according to the lattice parameter expansion obtained from the \gls{dft} calculations.
The source is a Gaussian pulse at normal incidence corresponding to the energy interval \unit[1--5]{eV}.
To mimic an infinite system, a \unit[100]{nm} thick \gls{pml} encloses the cell at a distance of at least \unit[100]{nm} from the nanodisk in all directions, to enable a sufficiently large vacuum region.
The cell is described in Cartesian coordinates with a grid resolution of \unit[0.5]{pixels/nm}.

A detailed description of the procedure of calculating extinction spectra can be found in \autoref{snote:extinction}.
In the data analysis, a peak in the extinction spectrum is defined as a point in the extinction spectra with larger amplitude than the two neighbouring points.
For the peak shift sensitivity calculation, only peaks with higher amplitude than the mean over the entire spectrum were included.
Calculating the sensitivity involves a linear fit of the studied shift with H content.
In this work, we employ Huber regression as implemented in \textsc{scikit-learn} \cite{PedVarGra11} for this purpose.

\subsection*{T-matrix simulations}\label{sect:tmat}
Nanodisk array extinction spectra were calculated using a T-matrix based approach as implemented in \textsc{SMUTHI} \cite{EGEL2021107846}.
The method allows for calculating the response of a substrate-supported single disk of varying geometry, as well as large arrays of composed of hundreds of such particles.
Note that for the nanodisk arrays, cylindrical disks where simulated instead of truncated cones, which has a negligible effect on the resulting extinction spectra (\autoref{sfig:fdtd-shape-effects}).
Arrays were treated either as amorphous and described statistically by the pair correlation function dependent on a  minimum center-to-center distance between disks (as previously demonstrated \cite{PhysRevLett.109.247401, PhysRevB.102.085431}) or as arrays consisting of 100 randomly distributed resonators with explicitly defined positions (see \autoref{sfig:array_res}b). For the latter case, size distribution within the array is also taken into account.

\section*{Supporting Information}

Supporting Information is available, free of charge, \textit{via} the \href{https://pubs.acs.org/}{ACS Publications website} and include: Additional visualizations of the \gls{dos} and \glspl{df}, nanodisk shape effects on the extinction spectrum, additional visualizations of extinction spectra, extended analysis of the double peak phenomenon of PdH, near-field enhancement obtained from FDTD simulations, hydrogen uptake as a function of the pressure and alloy composition, extended analysis of the peak shift sensitivity, effect of artificial broadening on the extinction spectra and the resulting sensitivity, particle size distribution and array effects on the optical response, optical response of the L1$_2$-ordered phase, visual representation of the \gls{df} fitting procedure.

Data pertaining to this article, including dielectric functions and extinction spectra, are provided via Zenodo at \url{https://doi.org/10.5281/zenodo.5833929}.

\begin{acknowledgments}
This work was funded by the Knut and Alice Wallenberg Foundation (2015.0055, 2019.0140), the Swedish Foundation for Strategic Research Materials framework (RMA15-0052), the Swedish National Research Council (2018-06482, 2020-04935), and the Excellence Initiative Nano at Chalmers.
T.P.R.\ acknowledges funding from Academy of Finland under grant agreement no.~332429.
T.J.A.\ acknowledges support from the National Science Center, Poland via the project 2019/35/B/ST5/02477.
The computations were enabled by resources provided by the Swedish National Infrastructure for Computing (SNIC) at NSC, C3SE and PDC partially funded by the Swedish Research Council through grant agreement no. 2018-05973 and the Interdisciplinary Center for Mathematical and Computational Modelling  via the project~\#GC84-51.
\end{acknowledgments}

\end{document}